\documentclass[twocolumn,showpacs,preprintnumbers,amsmath,amssymb,prl]{revtex4}

\usepackage{graphics}
\usepackage{graphicx}
\usepackage{amsmath,amssymb,epsfig,picinpar,graphics,float,amssymb,verbatim}

\newcommand{\nsum}[1]{\langle\,{#1}\,\rangle}
\newcommand{\ma}[1]{\mathcal{#1}}

\begin{document}

\title{Noise-induced macroscopic bifurcations in populations of
 globally coupled maps}
\author{Silvia De Monte$^*$, Francesco d'Ovidio$^{*,\dag}$, Erik Mosekilde$^*$}
\affiliation{Chaos Group$^*$ and Center for Quantum Protein$^\dag$, Dept. of
Physics,\\ Technical University of Denmark, DK 2800 Lyngby,
Denmark}
\date{\today}

\begin{abstract}
Populations of globally coupled identical maps subject to additive,
independent noise are studied in the regimes of strong
coupling. Contrary to each noisy population element, the mean
field dynamics undergoes qualitative changes when the noise strength
is varied. In the limit of infinite population size, these macroscopic
bifurcations can be accounted for by a deterministic system, where the
mean-field, having the same dynamics of each uncoupled element, is
coupled with other order parameters. Different approximation schemes
are proposed for polynomial and exponential functions and their
validity discussed for logistic and excitable maps.   
\end{abstract}

\maketitle

{\it Introduction.}
A fundamental question that arises when physical and biological populations 
are 
described by
means of mathematical models is how the collective dynamics is qualitatively 
affected
by random fluctuations acting at the microscopic level. While for systems at 
equilibrium a well
established thermodynamic theory exists, for out-of-equilibrium
phenomena, such as those typically described by nonlinear dynamical
systems, an overall theoretical view is still lacking.   
In this Letter we will address one aspect of this issue, namely the
effect of microscopic noise on the mean-field dynamics of large
populations of globally and strongly coupled identical maps.\\
Systems of globally coupled units are meant to model, among others,
Josephson junction
arrays \cite{josephson1},  yeast cells in a continuous-flow, stirred tank
reactor \cite{dano99} and neural cells \cite{sompolinsky91}, and
provide a mean field description of spatially extended systems with
long enough correlation length.\\
The effect of noise on one dynamical system has been extensively studied, 
pointing out phenomena such as noise-induced bifurcations, stochastic and 
coherence resonance (for a review, see Ref. \cite{toral01} and references 
therein).
Synchronization phenomena induced by (common or independent) noise have been 
studied for a system of two coupled dynamical systems 
\cite{maritan94, pikovsky94b, neiman95, andrade00, zhou02a, zhou02b}.
In the context of populations, the effect of microscopic noise on the 
collective dynamics has been recently inquired for continuous-time systems, 
namely phase models \cite{kurrer95, moro98, hong00, hasegawa02}, 
integrate-and-fire neurons \cite{rappel96, teramae02}, excitable systems 
\cite{sosnovtseva01,naundorf02}, chaotic systems \cite{zanette00, teramae01}. 
The onset of collective oscillations and the dependence of their frequency 
from the noise intensity has been mainly addressed.  
Analogous synchronization phenomena have been detected for stochastic 
oscillators \cite{nikitin01, naundorf02} and in spatially extended systems 
\cite{neiman99, ibanes01}.\\
Concerning maps, the relationship between single element and mean field 
fluctuations close to bifurcations \cite{nichols94} and the anomalous scaling 
of the population moments close to the onset of synchronization 
\cite{teramae01}  have been inquired.\\
However, the qualitative changes of the \emph{macroscopic dynamics} under the 
influence of noise have, at our best knowledge, never been systematically 
addressed. In particular, the strong coupling regimes could be considered of 
little interest since one expects the addition of a weak noise term not to 
greatly alter the mean field behavior.  Instead, in the first part of this 
Letter we give numerical evidence of
the fact that noise can qualitatively change the macroscopic
dynamics of the system: even if the single map has a noisy temporal
series, the mean field displays a low dimensional behaviour which can
be different with respect to the uncoupled map dynamics. 
The phenomenon of noise-induced macroscopic bifurcations will be
illustrated for two kinds of maps: logistic maps in the chaotic regime
and ``excitable maps''. 
Contrary to noise-induced synchronization, in this case it is the coupling 
that induces a coherent behavior for low noise intensity. The deterministic 
trajectory of each dynamical system is blurred out, so that the bifurcations 
detectable at a macroscopic level are a purely collective effect and
cannot be inferred by looking at one or few population elements. 
 
In the second part of the Letter, we explain these phenomena by
means of an order parameter expansion, valid for sufficiently large
coupling strength and large population size. This provides an approximate
description of the mean field dynamics in terms of few effective
macroscopic variables, whose \emph{deterministic} equations of motion
account for the macroscopic dynamics of the population and for the
bifurcations among different collective regimes.
The validity of such approach will be demonstrated on the two
aforementioned populations of maps. In particular, for the logistic
maps the bifurcation diagram of the
mean field will be rescaled to that of a single logistic map.

\begin{figure}[h]
\center
\epsfig{file=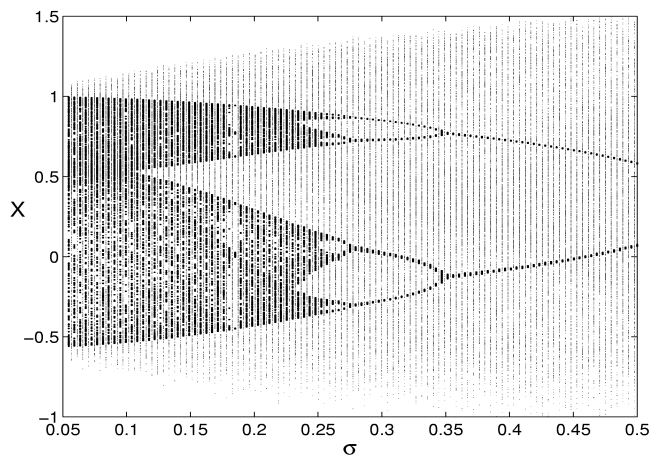, width=.48\textwidth,height=.34\textwidth}
\caption{Phase portrait of the mean field $X$ (big dots) of
  $N=2^{19}$ logistic 
  maps of the form $f(x)=1-a\,x^2$ ($a=1.57$, $k=0.7$) (and
  of one population element $x$ (small dots)) as a
  function of the noise standard deviation $\sigma$. The noise terms
  are generated according to a uniform distribution. \label{fig:log}}
\end{figure}

{\it Noise effect on coherent regimes.}
Let us consider populations of noisy and globally coupled identical
one-dimensional maps. We choose this system because it is sufficiently
simple to be analytically treated, and, at the same time, can be
considered as a prototype for inquiring new phenomena.\\
The fact that the mean field can exhibit complex dynamics is well known 
\cite{kaneko89,kaneko90,pikovsky94a,shimada02} and in particular we will focus 
on the synchronous regimes appearing for sufficiently strong coupling.\\
The equation of each population element is defined as follows:
\begin{equation}\label{eq:maps} 
x_j\mapsto (1-k)\:f(x_j)+k\:\nsum{f(x)}+ \xi_j(t) \qquad j=1,2,\dots
N,
\end{equation}
where $x_j\in \mathbb R$ is the state of the j-th population
element, $f:\mathbb R\rightarrow \mathbb R$ is a smooth function
defining the uncoupled elements dynamics and
$\nsum{f(x)}:=\sum_{j=1}^{N}\,f(x_j)/N$ is the average over 
the population. Every map is subject to a noise 
$\xi_j(t)$, that is determined according to a distribution of assigned
moments (in particular, it does not need to be Gaussian). The noise terms are 
independent and delta correlated, that is:
\[\xi_i(t)\xi_j(t')=\sigma^2\,\delta_{i,j}\,\delta_{t,t'},\]
where $\sigma^2$ is the variance of the noise terms distribution,
measuring the intensity of the microscopic noise.\\
Keeping all the parameter fixed, we will study the asymptotic
behaviour of the mean field
$X=\nsum{x}$ of large populations when $\sigma$, 
that will be from now on our control parameter, is changed.\\
Figure\ \ref{fig:log} shows the bifurcation diagram of $X$ in the case
of half a million logistic maps in the chaotic regime, together with the
phase portrait of an individual element of the population.
When the noise intensity is small, all the elements of the population
evolve coherently on a chaotic attractor. In the limit of zero noise,
indeed, all the elements are perfectly synchronized and have a common
chaotic trajectory (the assumption of strong coupling prevents the formation 
of clusters).
For larger noise intensities, the mean field displays an inverse
bifurcation cascade, crossing several periodic windows and undergoing
a period-halving scenario.
This simplification of the mean field dynamics does not however
reflect on the individual elements of the population, whose dynamics
is more and more smeared out with the increase of the microscopic
noise.\\
As we will later phrase in more rigorous terms, the fact that the
average has a regular 
behavior is a consequence of the fact that, in the population, 
a large number of simultaneous realizations of the noise occurs,
so that only statistical averages of the microscopic stochastic
process are relevant.\\
As a second example of bifurcations induced by microscopic noise we
consider an ``excitable map'' of the form (for other definitions of 
``excitable map'' see Refs. \cite{hayakawa00, toral01}):
\begin{equation}\label{eq:exc}f(x)=(\alpha\,x+\gamma\,x^3)\:e^{-\beta\,x^2}.
\end{equation} 
The parameters are
chosen in a region where the origin is the only fixed point, but, if the
system is initiated far enough from it, there is a chaotic transient. The
effect of noise on the single map is thus of exciting it
over threshold, so that a complex dynamics takes place.
\begin{figure}[h]
\center
\epsfig{file=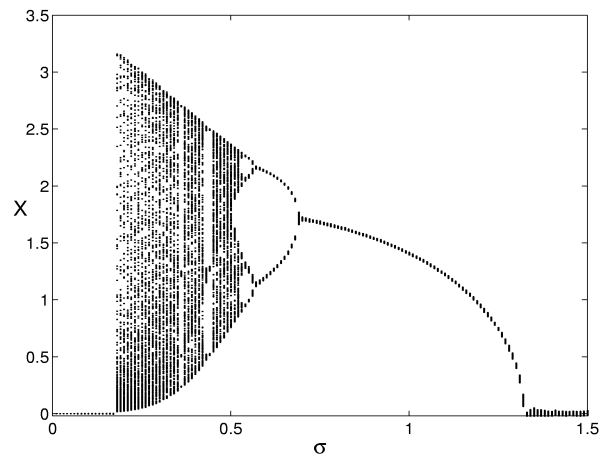, width=.48\textwidth,height=.34\textwidth}
\caption{Phase portrait of the mean field $X$ of
  $N=524288$ excitable maps of Eq.\ (\ref{eq:exc}) ($\alpha=.4$,
  $\beta=1$, $\gamma=8$, $k=0.9$) as a
  function of the noise standard deviation $\sigma$. The noise terms
  are generated according to a Gaussian distribution. \label{fig:exc}}
\end{figure} 
Correspondingly, the mean field has, for small noise intensity,
fluctuations above zero scaling as $\sigma/\sqrt{N}$, up to a critical
point, where the mean field starts displaying large
amplitude chaotic oscillations (Figure\ \ref{fig:exc}).  The
asymptotic dynamics simplifies 
for higher values of $\sigma$, the mean field going through a backwards
bifurcation cascade up to a steady state. For very large noise values,
the fixed point drops again to zero, as a
consequence of the fact that the map in Eq.\ (\ref{eq:exc}) is odd and
a strong noise causes the population to spread symmetrically around
the mean field.

{\it Order parameter expansion.}
Let us now address the problem from a mathematical viewpoint,
performing a change of variables that allows us to decouple the
macroscopic effect of noise from the dynamic ``skeleton'' furnished by
the uncoupled map equation.
This is achieved expressing the position of each population element in
terms of the mean field and of its displacement from it:
\begin{equation}\label{eq:varch}
x_j=X+\epsilon_j\hspace{20mm}j=1,2,\dots N.
\end{equation}
We can now substitute Eq.\ (\ref{eq:varch}) into the uncoupled element
equation and expand it in series around the mean field, thus obtaining:
\begin{equation}\label{eq:exp}
f(x_j)=f(X)+\sum_{q=1}^\infty\:\frac{1}{q!}\:\ma{D}^qf(X)\:\epsilon_j^q,
\end{equation}
where $\ma{D}^qf$ indicates the $q$-th derivative of the function $f$.
The equation for the mean field can now be obtained directly from its
definition, thus getting:
\begin{equation}\label{eq:X}
X\mapsto \nsum{f(x)}=
f(X)+\sum_{q=1}^\infty\:\frac{1}{q!}\:\ma{D}^qf(X)\:\nsum{\epsilon^q}
\end{equation}
We have at this point discarded the term $\nsum{\xi}$ that vanishes in
the limit of infinite population size and, otherwise, plays the role
of a macroscopic noise acting on the mean field and scaling as
$1/\sqrt{N}$, according to the law of large numbers.\\
Let us now define a set of new order parameters:
\begin{equation}\label{eq:ordpar}
\Omega_q:=\nsum{\epsilon^q}\hspace{20mm}q\in\mathbb N,
\end{equation}
and compute their evolution by making use of:
\begin{equation*}
\epsilon_j\mapsto(1-k)\:\sum_{p=1}^\infty\:\frac{1}{p!}
\:\ma{D}^pf(X)\:\left(\epsilon_j^p 
-\Omega_p\right)+\xi_j 
\end{equation*}
and of the fact that, being the positions and the noise uncorrelated
variables, in the limit $N\to \infty$ holds:
$
\nsum{h(X,\epsilon)\:\xi^q}=\nsum{h(X,\epsilon)}\:\nsum{\xi^q}.
$
From simple algebra thus follows:
\begin{eqnarray}\label{eq:omega}
\Omega_q\mapsto && m_q+
\sum_{i=1}^q \binom{q}{i} (1-k)^i\:m_{q-i}\nonumber\\
 &&\qquad\nsum{\left[\sum_{p=1}^\infty\:\frac{1}{p!}\:\ma{D}^pf(X)\:
\left(\epsilon^p-\Omega_p\right)\right]^i},
\end{eqnarray}
where $m_q=\nsum{\xi^q}$ is the $q$-th moment of the noise
distribution.
\begin{figure}[h]
\begin{minipage}{.5\textwidth}\centering
\epsfig{file=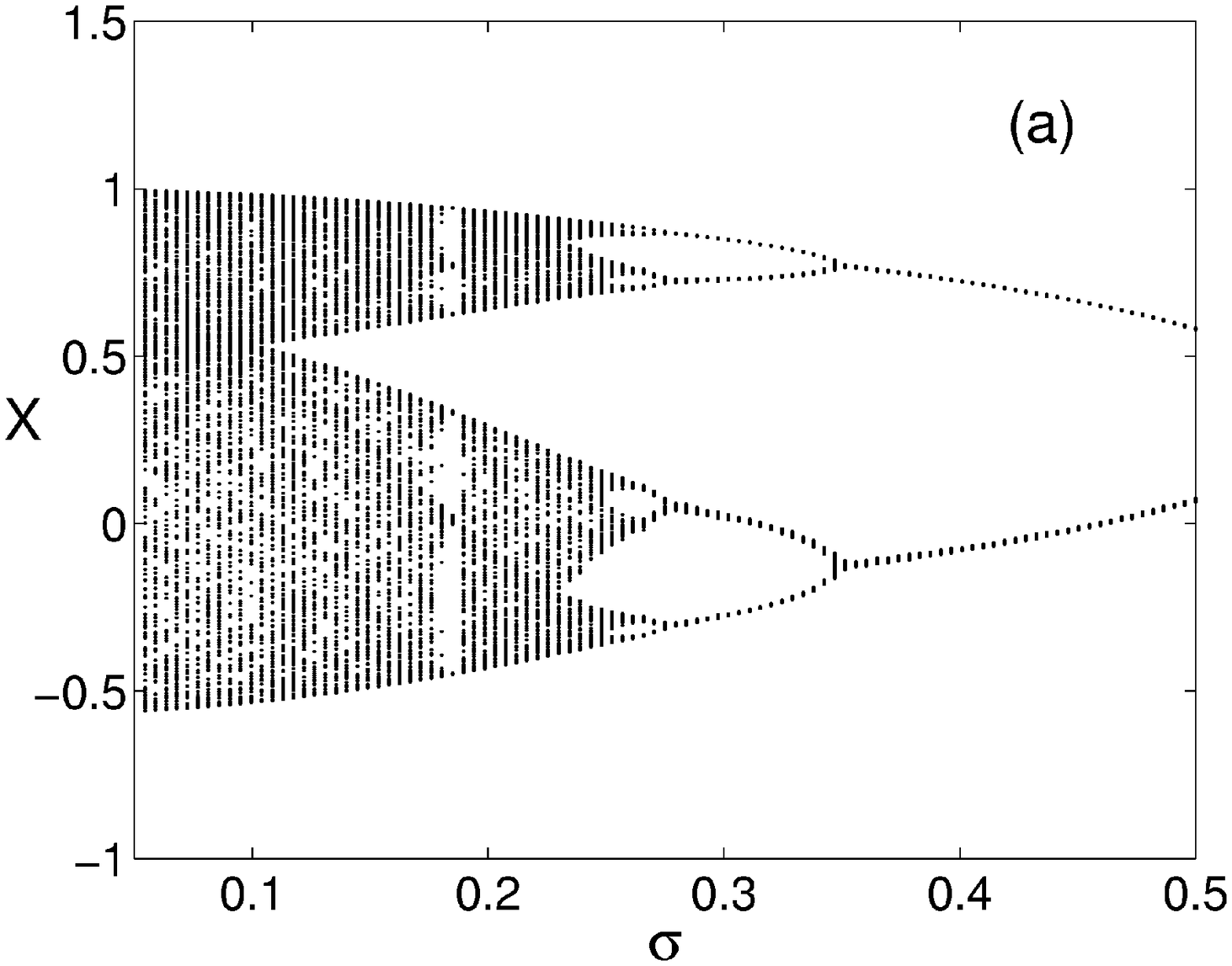, width=.48\textwidth,height=.35\textwidth}
\epsfig{file=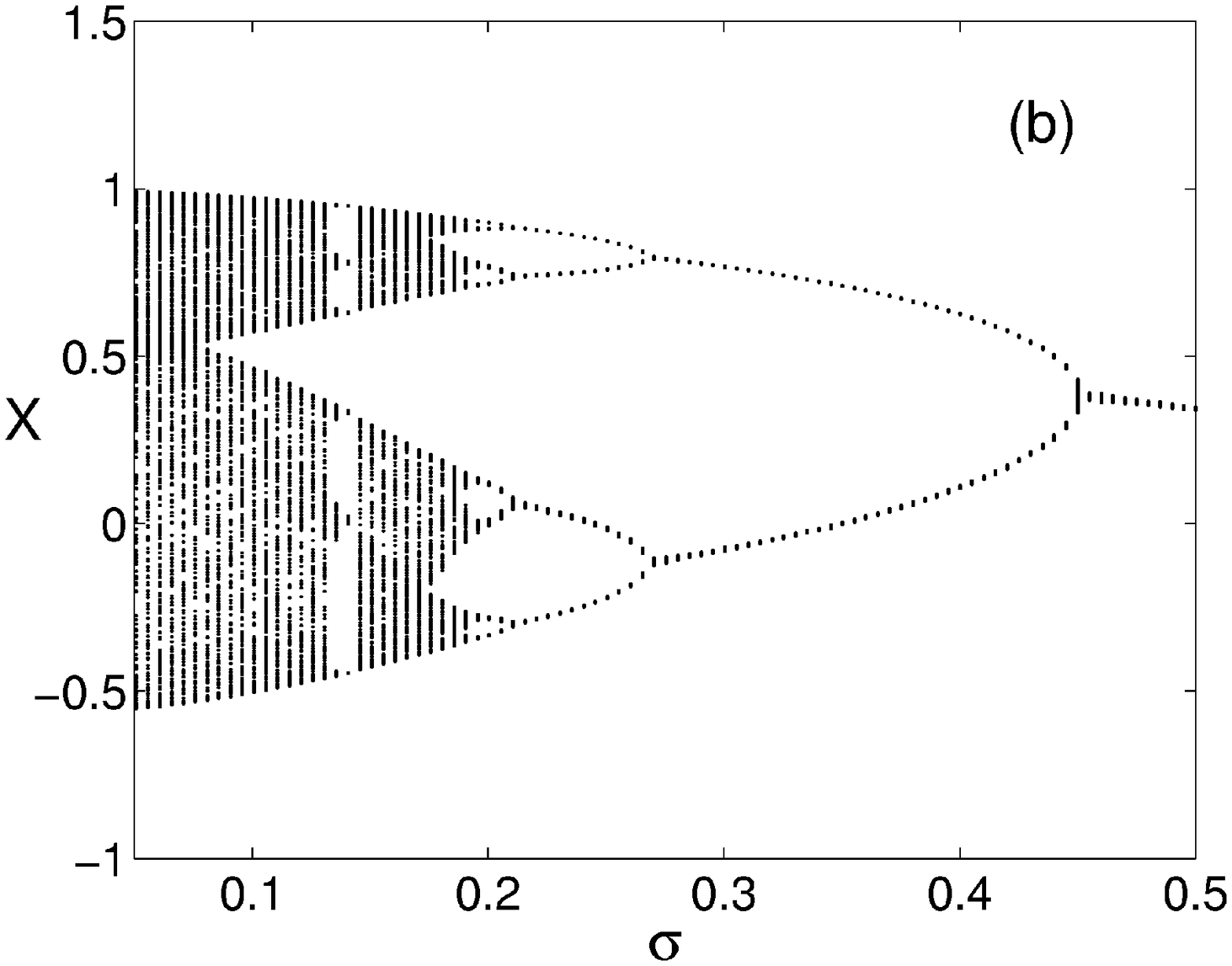, width=.48\textwidth,height=.35\textwidth}
\epsfig{file=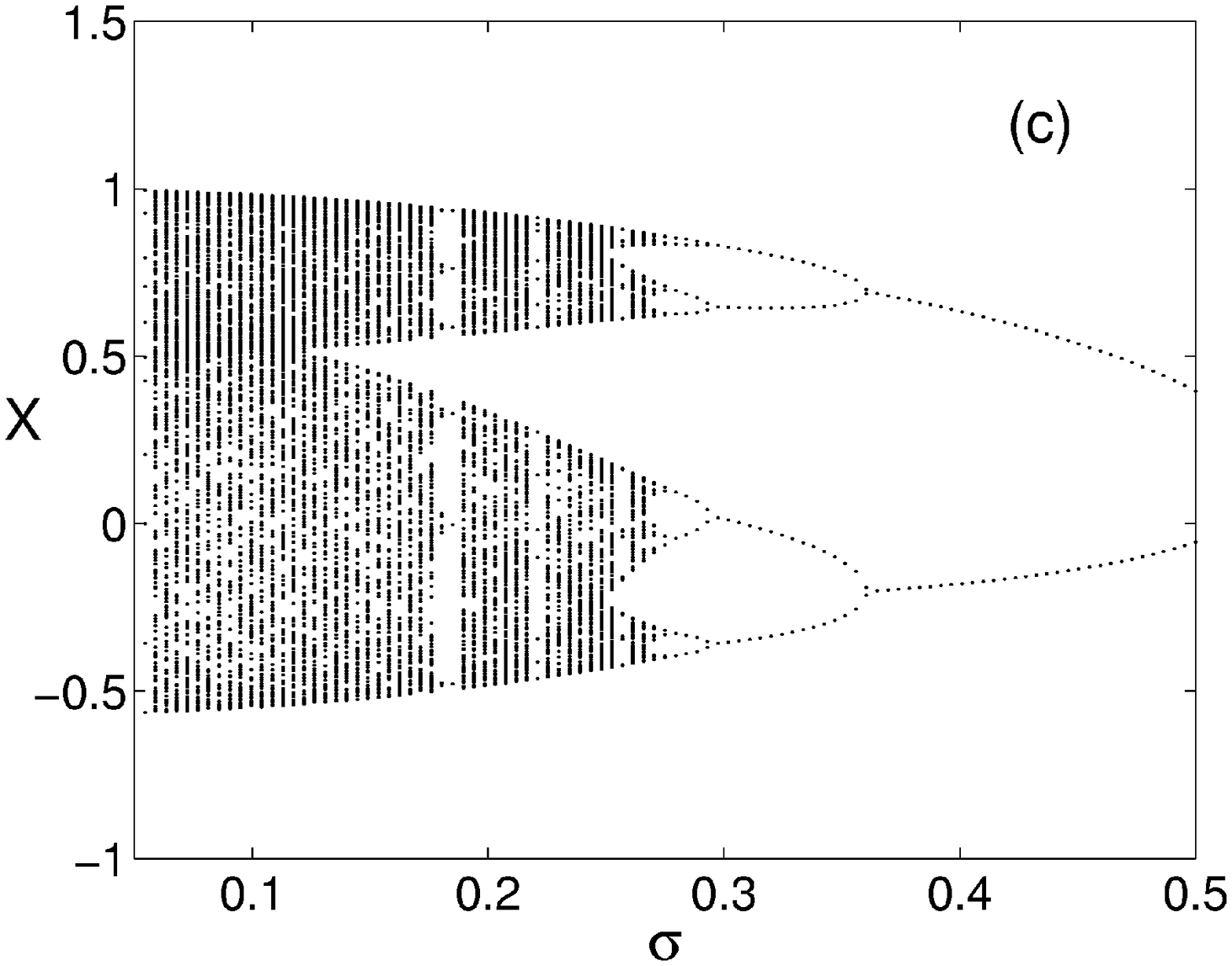, width=.48\textwidth,height=.35\textwidth}
\epsfig{file=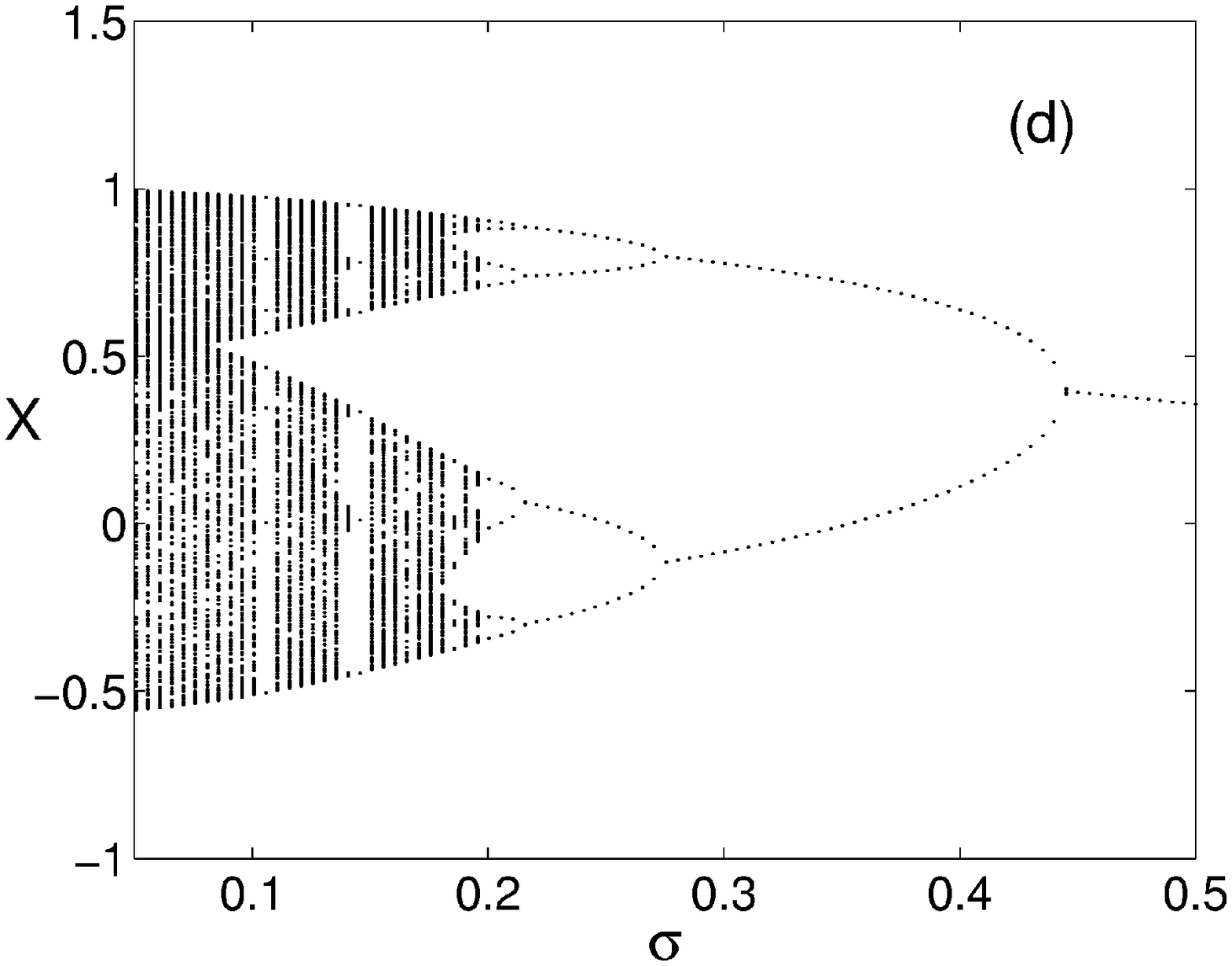, width=.48\textwidth,height=.35\textwidth}
\caption{Phase portrait, as a function of the (uniformly distributed)
  noise standard deviation  $\sigma$, of a) the mean field of a
  population ($N=524288$) of logistic maps of the form $f(x)=1-a\,x^2$
  ($a=1.57$) for high coupling ($k=0.9$); b) the mean field of the same
  population for weaker coupling ($k=0.7$); c) the zeroth-order
  expansion Eq.\ (\ref{eq:X0}); the second order expansion Eq.\
  (\ref{eq:exp2}). For lower coupling values, the qualitative agreement
  improves if further order parameters are taken into
  account.\label{fig:logcfr} }
\end{minipage} 
\end{figure}

\begin{figure}[h]
\begin{minipage}{.5\textwidth}\centering
\epsfig{file=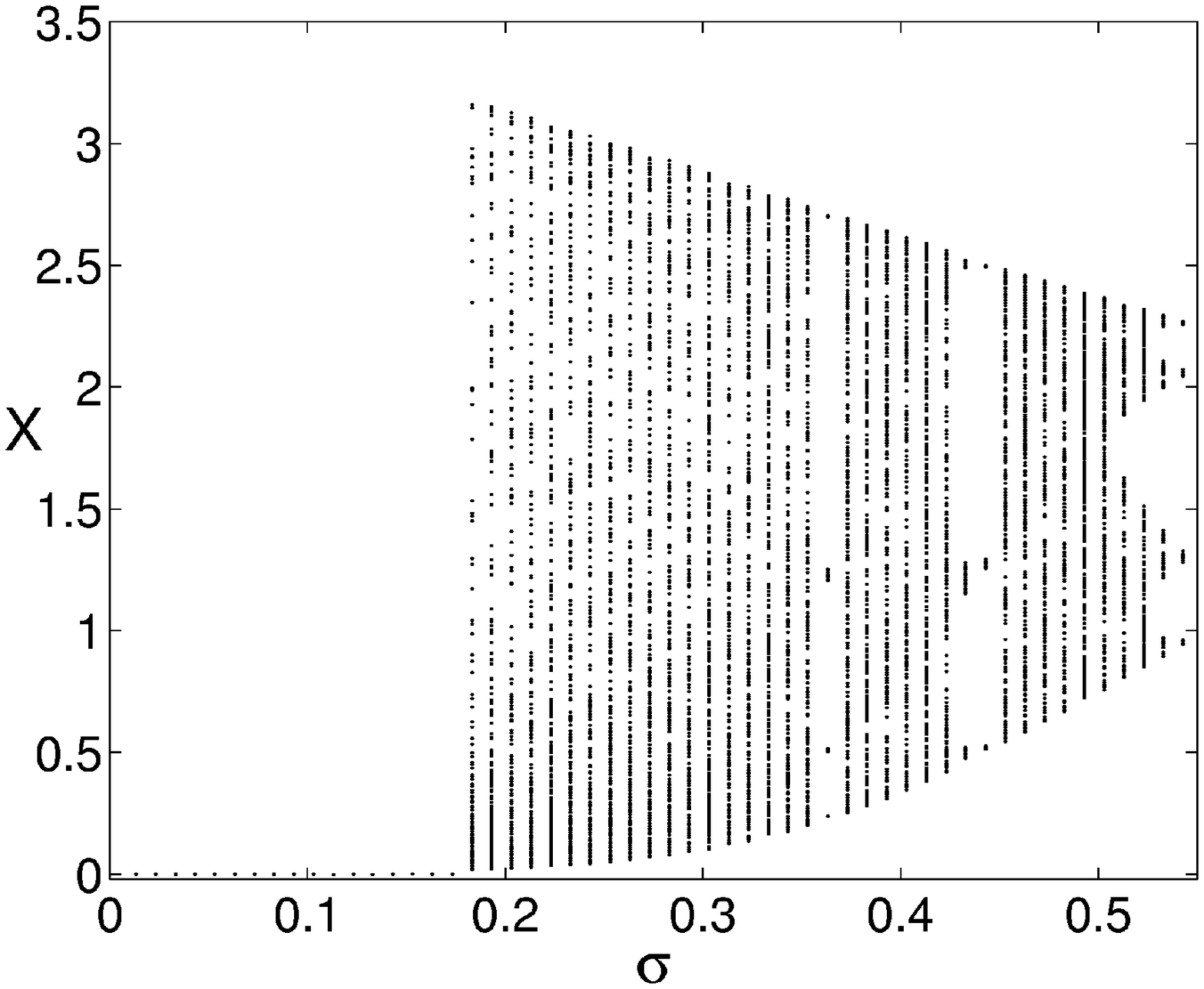, width=.48\textwidth,height=.35\textwidth}
\epsfig{file=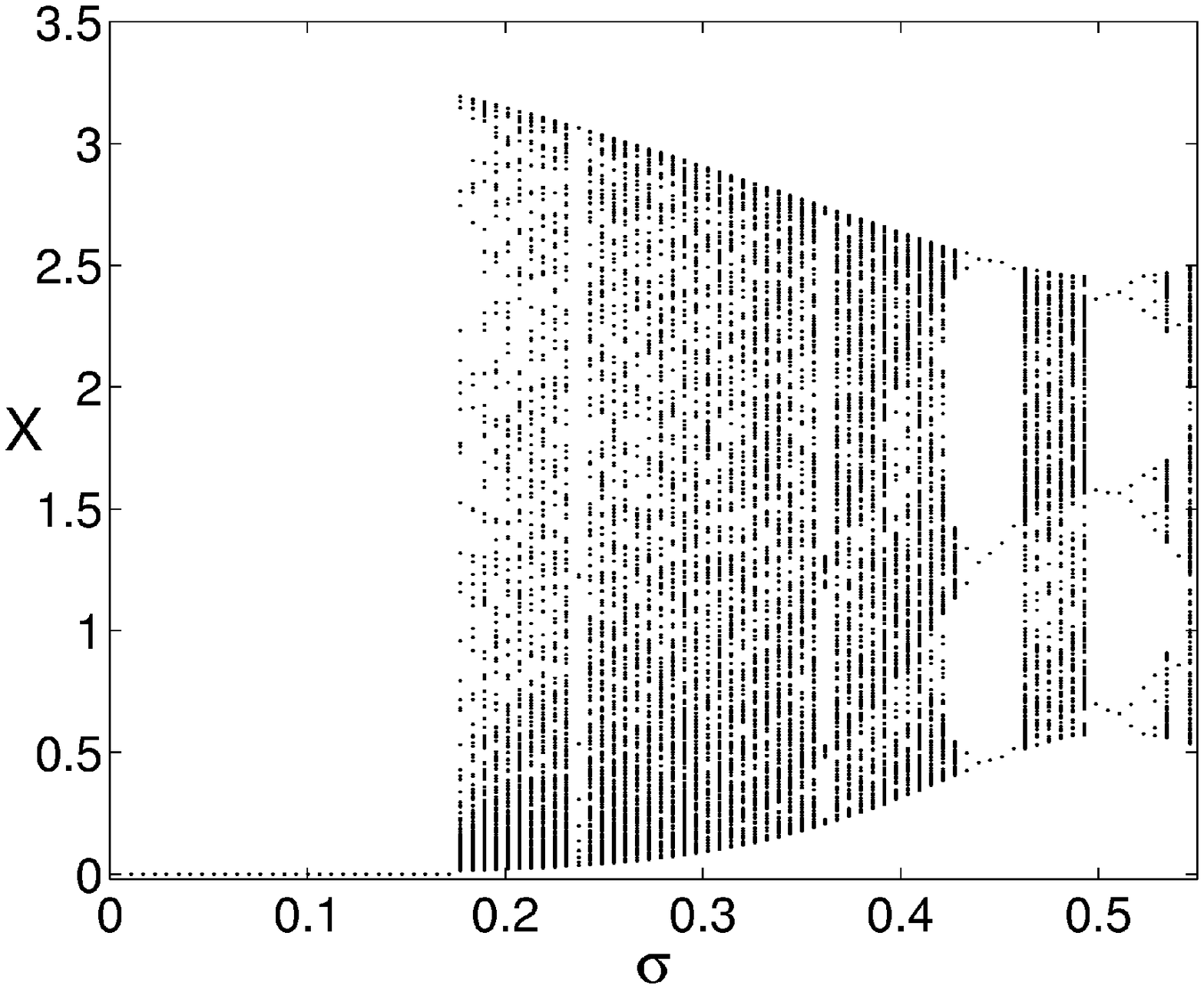, width=.48\textwidth,height=.35\textwidth}
\caption{Phase portrait as a
  function of the (normally distributed) noise standard deviation
  $\sigma$ of the mean field of a population ($N=2^19$) of excitable maps
  described by Eq.\ (\ref{eq:exc}) ($\alpha=6$,
  $\beta=1$, $\gamma=8$) for $k=0.9$ (top left) and for Eq.\ (\ref{eq:X0}) 
truncated at the sixth order. Lower order truncations lead to less accurate 
approximations (not shown).
  \label{fig:exccfr}} 
\end{minipage} 
\end{figure}

As a first approximation, valid for high coupling strength, the mean
field dynamics can be described by a scalar equation:
\begin{equation}\label{eq:X0}
X\mapsto 
f(X)+\sum_{q=1}^\infty\:\frac{1}{q!}\:\ma{D}^qf(X)\:m_q
\end{equation} 
which is obtained by Eqs.\ (\ref{eq:X}) and (\ref{eq:omega}) in the
limit $k\to 1$.\\
The two left-hand pictures of Figure\ \ref{fig:logcfr} show the mean
field dynamics 
for the population of logistic maps under high coupling
($k=0.9$, top) and for the effective mean field 
dynamics Eq.\ (\ref{eq:X0}) (bottom), that in this case takes the simple form:
\begin{equation}\label{eq:log0}
X\mapsto 1-a\:\sigma^2-a\:X^2.
\end{equation}  
This equation provides a rescaling of the average population dynamics
to that of a single logistic map of the form $b+a\:X^2$.\\
A more difficult case to analyse is that of the excitable maps, since
in this case the sum in Eq.\ (\ref{eq:X0}) contains an
infinite number of terms. 
Any finite truncation thus introduces a further approximation, whose validity 
holds for sufficiently weak noise. Figure\ \ref{fig:exccfr} reports the mean 
field bifurcation
diagram for the population (left) and for the sixth-order 
truncation of Eq.\ (\ref{eq:X0}) (right), that is able to describe the onset 
of the collective chaotic oscillations up to the first periodic windows. 
It is worth noticing that the terms of high order in the series of 
Eq.\ (\ref{eq:X0}) are small for low noise intensity, so being the
noise distribution moments. Therefore, increasing the noise variance
from zero causes the nonlinearities of the uncoupled map to become
progressively important in the mean field dynamics description.
In general, we can see that Eq.\ (\ref{eq:X0}) accounts for the
influence of the noise distribution features (described by its
moments) on the dynamics of an uncoupled element. If this is a
polynomial, the $q$-th moment will affect the coefficients of the
terms of order less than $q$. It is however important to stress that
no requirements have been made on the specific shape of the noise
distribution, so that distributions having the same low-order moments
have similar macroscopic effects for weak noise. 

Let us now consider the case of lower coupling and include in the
effective description also the terms in $(1-k)^2$ (those in $(1-k)$
cancel out for symmetrical noise distributions, like those considered so far):
\begin{eqnarray}\label{eq:exp2}
X\mapsto&& 
f(X)+\sum_{q=1}^\infty\:\frac{1}{q!}\:\ma{D}^qf(X)\:\Omega_q\nonumber\\
\Omega_q\mapsto&& m_q+ \frac{q(q-1)}{2}\:(1-k)^2\:m_{q-2}\\
&&\qquad \sum_{p=1}^\infty \frac {1}{p^2!}
\:\left[\ma{D}^p\,f(X)\right]^2 \:\left(\Omega_{2p}-\Omega_p^2\right)\nonumber
\end{eqnarray}  
If $f$ is a polynomial of order $n$, the reduced system Eq.\
(\ref{eq:exp2}) is a closed system of $2n+1$ variables, which thus
determine the dimensionality of the mean field dynamics.
Again, when $f$ is not a polynomial, any truncation of Eq.\
(\ref{eq:exp2}) introduces errors which become bigger when the noise
intensity increases. Nevertheless, the moments of 
high order become negligible when $\sigma$ is small and the mean field
dynamics can still be described in the weak noise regimes. 
\\
Going back to the previously considered population of logistic
maps, we lower the
coupling strength. Equation\ (\ref{eq:exp2}) takes now the following form:
\begin{eqnarray}\label{eq:log2}
X\mapsto&& 1-a\,X^2-a\:\Omega_2\nonumber\\
\Omega_2\mapsto&& \sigma^2+(1-k)^2\,a^2
\left(4\,X^2\,\Omega_2-\Omega_2^2+\Omega_4\right)\\
\Omega_4\mapsto&& m_4+6(1-k)^2a^2\sigma^2
\left(4X^2\Omega_2-\Omega_2^2+\Omega_4\right)\nonumber
\end{eqnarray}
The two right-hand images of Figure\ \ref{fig:logcfr} compare the mean
field dynamics for $k=0.7$ (top) 
with that of the second order approximation Eq.\ (\ref{eq:log2}) (bottom). the
mean field dynamics can be directly compared with the zeroth-order
approximation (bottom left of the same figure), that is independent of
$k$.
It is evident that the accordance with the population bifurcation
scenario is improved when more order parameters are introduced in the
description. In particular, Eq.\ (\ref{eq:log2}) reproduces the shift of the 
period-doubling cascade toward lower values of the noise intensity, taking 
place when the coupling is weakened.

{\it Conclusions.}
In this Letter we have discussed the phenomenon of macroscopic
bifurcations induced by microscopic additive noise in large populations of
globally and strongly coupled maps.
We have shown that the macroscopic dynamics changes with the noise
intensity and that in the case of high coupling the macroscopic
bifurcations can be reproduced by a low-dimensional map, where the
mean field is coupled to some additional order parameters. This order
parameter reduction accounts for the effects of noise on the dynamical
``skeleton'' given by the uncoupled element equation.
The proposed method holds for any smooth map and is
largely independent from the 
specific characteristics of the noise distribution, allowing to
understand how the noise moments interact with the nonlinearities of the
uncoupled map equation. 
As examples of application, we have considered logistic maps in the
chaotic regime and excitable maps, and we have addressed the validity
of different approximation schemes.\\
Although populations of scalar maps have been considered in order to
simplify both the analytical and the numerical work, studies on 
continuous-time systems (in particular, Refs. \cite{kurrer95,hong00} together 
with our
preliminary results indicate that microscopic noise might have a
similar effect on strongly coupled continuous time systems, inducing
macroscopic bifurcations of the mean field.
The method presented here could be combined with the order parameter
expansion proposed in Ref. \cite{demonte03} for investigating the different 
roles of
microscopic noise and intrinsic parameter diversity on populations of
dynamical systems.

\bibliography{noisebib}

\end{document}